\documentclass[useAMS,usenatbib]{mn2e}

\usepackage{graphicx}
\usepackage{epsfig}
\usepackage{amsmath}
\usepackage{amssymb}
\usepackage{subfigure}
\usepackage{epsfig}
\usepackage{array}
\usepackage{lscape}
\usepackage{color}
\usepackage{longtable}
\usepackage{arydshln}

\def\nat{\reff@jnl{Nature}}  
\def\reff@jnl#1{{\rm#1\/}}
\def\aj{\reff@jnl{AJ}}                  
\def\araa{\reff@jnl{ARA\&A}}            
\def\apj{\reff@jnl{ApJ}}                
\def\apjl{\reff@jnl{ApJ}}               
\def\apjs{\reff@jnl{ApJS}}     
\def\aap{\reff@jnl{A\&A}}  
\def\mnras{\reff@jnl{MNRAS}}  
\def\pasj{\reff@jnl{PASJ}}              

\begin{document}

\newcommand{\cl}{Cl\,0016+16}
\newcommand{\four}{MS\,0451.6-0305}
\newcommand{\ten}{MS\,1054.4-0321}
\newcommand{\rx}{RXJ\,2129.6+0005}

\title[]{Sunyaev Zel'dovich observations of a statistically complete
sample of galaxy clusters with OCRA-p}

\author[]{Katy Lancaster$^1$, Mark Birkinshaw$^1$, Marcin
  P. Gawro\'nski$^2$, 
\newauthor Richard Battye$^3$, Ian Browne$^3$, Richard
  Davis$^3$, Paul Giles$^1$, Roman Feiler$^2$, 
\newauthor Andrzej Kus$^2$, Bartosz Lew$^2$, Stuart Lowe$^3$, Ben Maughan$^1$, 
Abdulaziz Mohammad$^1$,
\newauthor Bogna Pazderska$^2$, Eugeniusz Pazderski$^2$, Mike
  Peel$^3$, Boud Roukema$^2$
\newauthor and Peter Wilkinson$^3$\\ 
  $^1$ H. H. Wills Physics
  Laboratory, University of Bristol, Tyndall Avenue, Bristol BS8 1TL\\
  $^2$ Toru\'n Centre for Astronomy, Nicolaus Copernicus University,
  ul. Gagarina 11, 87-100 Toru\'n, Poland\\ 
  $^3$ Jodrell Bank Centre for Astrophysics, The University of Manchester, 
  Alan Turing Building, Manchester M13 9PL\\ }

\date{Received **insert**; Accepted **insert**}

\pagerange{\pageref{firstpage}--\pageref{lastpage}} 
\pubyear{}

\maketitle
\label{firstpage}

\begin{abstract}

\noindent We present 30\,GHz Sunyaev Zel'dovich observations of a
statistically complete sample of galaxy clusters with OCRA-p.  The
clusters are the 18 most X--ray luminous clusters at $z > 0.2$ in the
ROSAT Brightest Cluster Sample.  We correct for contaminant radio
sources via supplementary observations with the Green Bank Telescope,
also at 30\,GHz, and remove a cluster that is contaminated by an
unresolved X-ray source.  All 17 remaining clusters have central SZ
effects with Comptonisation parameter $y_0$ exceeding
$1.9\times10^{-4}$, and 13 are detected at significance $\ge 3
\sigma$. We use our data to examine scalings between $y_0$ and X-ray
temperature, X-ray luminosity, and the X-ray mass proxy
$Y_\mathrm{X}$, and find good agreement with predictions from
self--similar models of cluster formation, with an intrinsic scatter
in $y_0$ of about 25\%.  We also comment on the success of the
observations in the face of the contaminant source population, and the
implications for upcoming cm-wave surveys.

\end{abstract}



\section{INTRODUCTION}
\label{intro}

The thermal Sunyaev Zel'dovich (SZ) effect (\citealp{SZ}) is a
spectral distortion of the Cosmic Microwave Background (CMB) radiation
due to inverse Compton scattering by the hot gas in galaxy clusters.
It has long been exploited in cosmological and cluster studies in
order to derive, for example, the Hubble constant
(e.g. \citealp{Hughes1998}, \citealp{Mason_obs}, \citealp{Reese2002},
\citealp{Saunders2003}, \citealp{Bonamente2006}) and the gas mass
fraction (e.g. \citealp{Grego2001}, \citealp{Lancaster2005},
\citealp{LaRoque2006}).  Thanks to well developed techniques,
detections are becoming routine although signal--to--noise remains
quite poor.  However, we are entering an era of purpose--built
instruments so this is set to improve dramatically, enabling SZ
research to reach its evident potential.

The main focus of the SZ community at present is to utilise the
redshift--independence of the SZ surface brightness in order to
perform blind surveys for galaxy clusters.  While other techniques
suffer from large intrinsic biases and complex selection effects, SZ
surveys will produce almost \emph{mass--limited} catalogues and thus
far superior datasets for constraining cosmological models.  The
dedicated SZ surveys, for example Planck (\citealp{Ade2011}), SPT
(\citealp{Stan2009}, \citealp{Vanderlinde2010s}), ACT
(\citealp{Menanteau2010}, \citealp{Marriage2010}) and the SZA
\citep{Muchovej2011} are now generating results.  Many more are
expected in the near future, e.g. from AMI \citep{Zwart2008s}.  In
order to fully exploit the results of these surveys, it will be
necessary to improve understanding of both the `selection effect' due
to the presence of unsubtracted radio sources, and also the scalings
between cluster SZ observables and various physical quantities,
especially the cluster mass.  Various groups have undertaken similar
studies (e.g. \citealp{Benson2004}, \citealp{Morandi2007},
\citealp{Bonamente2008}, \citealp{Huang2010}).  To make further
advances, reliable observations of large, well--selected samples are
required.

The One Centimetre Receiver Array prototype, OCRA-p, is a two--element
receiver mounted on the 32-m telescope at the Toru\'n Centre for
Astronomy of the Nicolaus Copernicus University, Poland.  OCRA--p
proved its SZ capabilities by detecting four well known clusters at
high significance \citep{Lancaster2007}.  We present the first in a
series of papers discussing an X--ray selected sample of 33 clusters
with well understood selection effects.  This paper contains details
of our OCRA--p observations of a statistically complete subsample of
18 clusters.  The structure of the paper is as follows.  In Section 2
we give a brief description of the Toru\'n 32-m telescope and the
OCRA-p receiver.  Section 3 contains details of the cluster sample.
The observations and data reduction are described in Section 4, and
the problem of radio source contamination is detailed in Section 5.
Section 6 gives a brief overview of our X-ray analysis, and Section 7
presents our results. We discuss these results and conclude in Section
8. Throughout the paper we adopt the following cosmological parameter
values: $H_0 = 70 \, \mathrm{km} \, \mathrm{s}^{-1} \,
\mathrm{Mpc}^{-1}$, $\Omega_{\mathrm{m}0} = 0.3$, $\Omega_{\Lambda 0}
= 0.7$.


\section{THE TORUN TELESCOPE AND OCRA-p}

The Toru\'n observatory is located in Piwnice, 15\,km outside Toru\'n
in northern Poland.  The telescope consists of a 32-m parabolic dish
and 3.2-m sub-reflector, with a fully steerable classical Alt--Az
mount.  It has receivers operating at 1.4-1.7, 5, 6.8, 12 and 30\,GHz.
The telescope is used for a variety of studies including interstellar
molecular spectroscopy (e.g. \citealp{Blaszkiewicz2004}) and VLBI
(e.g. \citealp{Bartkiewicz2005}).

OCRA is a planned 30\,GHz 100--element continuum receiver (see
\citealp{Browne2000}).  A prototype receiver, OCRA-p, funded by a
grant from the Royal Society Paul Instrument fund, has been on the
telescope since 2004.  The instrument is described in detail in
\cite{LowePhD} and \cite{Lowe2007}; here we present a short summary.
The basic radiometer design is based on the prototype for the Planck
Low Frequency Instrument (LFI, \citealp{Mandolesi2000}), and is
similar to the WMAP K-band receivers \citep{Jarosik2003}.  OCRA-p
provides only two horn--feeds, the beams of which are separated by
$3'.1$ and have FWHM $1'.2$.  As the beam separation is small, it is
possible to reduce the effects of atmospheric and gain fluctuations by
switching between the beams and taking the difference.  This can be
improved upon by further levels of switching.  The full switching
strategy for SZ observations is described in Section \ref{sec:obs}.
In addition to the SZ program, OCRA--p has been used to study
planetary nebulae (\citealp{Pazderska2009}) and for various radio
source studies including flux density measurements to support the Very
Small Array CMB program (see \citealp{Gawronski2010} for details of
the OCRA measurements, and Genova--Santos et al. (in preparation) for
the details of the CMB work).  The next phase of the project is an
8--element array, OCRA--Faraday (\citealp{Peel2009}), which was
mounted on the telescope in late 2009.  This is still in the
commissioning phase but is expected to be fully operational in autumn
2011, and can be upgraded to 16 elements in the future.


\section{THE CLUSTER SAMPLE}
\label{sec:clusters}

In order to derive meaningful cosmological results, or indeed to
comment on universal cluster properties, it is necessary to use a
`fair' sample, or at least one where the selection biases are clear
and well understood.  Large samples fulfilling these criteria are rare
in SZ astronomy, largely due to the various practical challenges which
observers must overcome.  In this work, we attempt to select and observe such
a sample, and turn to the ROSAT All--Sky Survey in the first instance.

The Brightest Cluster Sample (BCS, \citealp{Ebeling1998}) is derived
from a careful analysis of the ROSAT All-Sky Survey (RASS,
\citealp{Voges1999}) data at $\delta > 0^{\circ}$.  The resulting
catalogue is 90\% flux complete to a limit of $f_{\mathrm{X}} = 4.4
\times 10^{-12}$\,erg\,cm$^{-2}$\,s$^{-1}$ in the 0.1-2.4\,keV energy
band.  Since the SZ surface brightness of a cluster is essentially
unaffected by redshift, we select on luminosity, imposing a limit of
$L_{\mathrm{X}} \geqslant 13.08\times10^{44}\mathrm{erg\,s^{-1}}$
(the luminosity of A773 in the BCS catalogue).  In addition,
we impose the criterion $z>0.2$ due to the $3'.1$ beam--throw of OCRA;
observations at lower redshift are inefficient with this
instrument.  Our basic sample then consists of 18 clusters.  We do not
reject any clusters based on their radio--source environments at this
stage, even though we appreciate the difficulties which this policy
may introduce.  This has the advantage of allowing us to investigate
any correlation between gas properties and those of a central radio
galaxy.

We reject one cluster from the basic 18--cluster sample, A689, as on
further investigation we have found that the ROSAT data are
contaminated by a bright central point source (Giles et al., in
preparation).  After removing this source from the \emph{Chandra}
data, the luminosity is just $(1.66\pm0.24)\times10^{44} \mathrm{erg
 \,s}^{-1}$, more than a factor 8 below the limiting luminosity of the
cluster sample.  The sample is summarised in Table \ref{tab:coord}.
These clusters appear not to be exceptional in their X--ray
structures, and, indeed, follow the expected X-ray scaling relations
(see Section \ref{sec:Xscale}).

\begin{table*}
\caption{Initial cluster parameters.  Coordinates correspond to the
  OCRA pointing centres which in some cases were shifted slightly to
  avoid bright point sources. $\Delta$RA and $\Delta$Dec are the
  offsets from the maxima of the X--ray surface brightness in the
  \emph{Chandra} data.  $L_\mathrm{X}$ values refer to the
  0.1-2.4\,keV ROSAT energy band and are taken from
  \protect\citet{Ebeling1998} along with the redshifts. A689 was
  originally in the sample, but on closer inspection it was found to
  have significant X-ray point source contamination.  Once this was
  removed, the cluster luminosity fell to the bracketed value, well
  below our selection limit of $13.02\times10^{44}$ erg\,s$^{-1}$.}

\begin{tabular}{lcccccc}
\hline Cluster &RA &Dec &$\Delta$RA &$\Delta$Dec &$z$ &$L_\mathrm{x}$\\
&(J2000) &(J2000) &(arcmin) &(arcmin) & &($10^{44}$erg\,s$^{-1}$)\\ \hline
A1835 &14 01 01.99 &02 53 12.8 &0.02 &0.53
&0.25 &38.53\\
ZWCL1953 &08 50 03.00 &36 04 16.0 &-0.80 &-0.04 
&0.32 &34.12 \\
\emph {A689} &\emph{08 37 25.01} &\emph{14 59 40.9} &\emph{-} &\emph{-} 
&\emph{0.28} &\emph{30.41 (1.66)}\\
ZWCL3146 &10 23 35.98 &04 11 56.0 &-0.90 &0.75
&0.29 &26.47 \\
RXJ1532.9+3021 &15 32 58.99 &30 21 11.2 &1.13 &0.20
&0.35 &24.40 \\
A2390 &21 53 34.61 &17 40 10.9 &-0.50 &-1.56
&0.23 &21.44 \\
A2219 &16 40 21.80 &46 42 47.8 &0.24 &0.30
&0.23 &20.40 \\
RXJ2129.6+0005 &21 29 40.50 &00 05 49.9 &0.11 &0.51
&0.24 &18.59 \\
A2261 &17 22 27.00 &32 07 04.1 &-0.03 &-0.88
&0.22 &18.18 \\
A781 &09 20 25.99 &30 29 57.8 &-0.04 &-0.11
&0.30 &17.22 \\
A697 &08 42 58.01 &36 21 45.0 &0.03 &-0.20
&0.28 &16.30 \\
A1763 &13 35 23.21 &41 00 04.0 &0.04 &0.08
&0.22 &14.93 \\
A68 &00 37 05.30 &09 09 10.8 &-0.20 &-0.38
&0.26 &14.89 \\
A520 &04 54 19.01 &02 56 48.1 &0.00 &0.00
&0.20 &14.44 \\
A267 &01 52 42.00 &01 00 25.9 &-0.02 &-0.24
&0.23 &13.71 \\
RXJ0439.0+0715 &04 39 01.01 &07 16 54.8 &0.08 &0.85
&0.23 &13.25 \\
ZWCL7160 &14 57 19.99 &22 20 35.9 &1.13 &0.03
&0.26 &13.19 \\
A773 &09 17 51.00 &51 43 19.9 &-0.32 &-0.34
&0.22 &13.08 \\
\hline
\end{tabular}
\label{tab:coord}
\end{table*}


\section{OCRA DATA}

\subsection{Observing Strategy}
\label{sec:obs}

The clusters were observed in good weather periods between September
2005 and January 2009.  We aimed to achieve a uniform noise level
$<0.5$\,mJy, which required $\sim 220$ minutes of data at each
pointing centre. Trail fields, blank patches of sky separated from the
target fields by 15--20m in RA, were observed over the same range of
hour angle after each cluster observation, using the same position
switching routine (described below) as for the cluster fields.  The
integration times for the target and trail fields are identical.

For each observation, whether the cluster or its corresponding trail
field, we employ a position switching strategy to remove atmospheric
contamination from our data.  The feeds are positioned such that one
beam, beam $B$, is coincident with the cluster centre and the other,
beam $A$, provides a measure of the blank sky signal.  (For extended
sources, beam $A$ may measure a small signal itself.  This must be
properly accounted for - see Section \ref{sec:beta}.) We switch
between the $A$ and $B$ beams at a rate of 277\,kHz, recording the
$A-B$ difference every second.  We integrate in this position for 25
seconds. The telescope then slews to reposition the beams such that
beam $A$ measures the cluster and beam $B$ measures the sky
background, and the differencing is repeated, this time integrating
for 50 seconds.  We then return to the starting position and integrate
for another 25 seconds.  The differenced signals from the first and
third positions, $(A_1-B_1)$ and $(A_3-B_3)$ are summed, and then
subtracted from the differenced signal from the second position
$(A_2-B_2)$.  This recovers twice the cluster signal relative to the
background regions. The symmetric nature of the switching pattern
ensures proper subtraction of most atmospheric emission and other
contaminants such as ground pick-up, provided that the target region
is not rapidly changing in azimuth and elevation.

\subsection{Calibration}

The data are calibrated against an internal noise source, which is
itself calibrated via observations of the well--known bright radio
source NGC7027 of flux density 5.37\,Jy.  We observe significant
changes in the level of the noise source in the form of: (a) sudden
`jumps', sometimes due to the telescope reconfiguration and (b)
additional smaller amplitude fluctuations which appear random in
nature.  We note that other than the low--amplitude random
fluctuations, the voltage level is generally stable on timescales of a
few days.  By performing regular calibration ($\approx$\,daily) we
reduce the effects of the larger fluctuations.  In addition, we adopt
a conservative approach by comparing the total powers for the NGC7027
and cluster observations and rejecting any data for which they differ
significantly, although a study suggests the calibration often stays
virtually the same as the total power drifts.  Residual
uncertainties in the calibration of the system at the level of 5\% may
be expected.

The telescope pointing is calibrated via azimuth and elevation scans
across a bright source close to each cluster.  Where an acceptable
pointing calibration does not exist within 60\,minutes of a cluster
observation, the cluster data are rejected.

\subsection{Statistical Data Analysis}

After combination of the second-by-second average data into
double--differenced measurements of the brightness of the sky, and
calibration, the data are examined for periods of increased noise
(which might arise from receiver instabilities or bad weather
conditions) or individual anomalous points. The combination of the
data into final averages was performed including statistical tests for
outlier data. The fractions of the data points rejected by $3\sigma$
or $5\sigma$ cuts were small in all cases, and no cut-dependent
changes larger than a small fraction of the error on the mean were
seen.  The distributions of data values in the double-difference data
are close to Gaussian, with a slight tendency to show elevated wings
in the distributions: the estimates of the error on the mean
(Table \ref{tab:res}) take account of the full distribution, and not the
distribution after truncation of outliers.


\section{RADIO SOURCE CONTAMINATION}
\label{sec:src}

Radio source contamination remains a significant problem for
observations of the Sunyaev Zel'dovich effect.  If we know the
positions and 30\,GHz flux densities of all sources near our observing
locations, we can correct the data and establish the true magnitude of
the SZ effect for each cluster.  Ideally, we would consult a survey of
the radio sky at a frequency close to 30\,GHz in order to identify
contaminant radio sources.  The only such survey currently available
is the WMAP point source catalogue (\citealp{Wright2009}), but this is
complete only to the 2\,Jy level and so is three orders of magnitude
too shallow for our purposes.  We consult the all--sky NVSS
(\citealp{NVSS}, 1.4\,GHz) and GB6 (\citealp{GB6}, 4.85\,GHz)
catalogues and retain all sources within $5'$ of the pointing centre
for each cluster.  In addition, we include any sources found in
similar projects and reported in the literature.  The list of 58
contaminating sources is presented in Table \ref{table:sources}.

\subsection{GBT observations}

We observed the 58 sources over the course of several sessions at the
 Green Bank Telescope (GBT) in January 2008 (observations limited due
 to poor weather conditions), then January and May 2009 (re-scheduled
 time, again compromised in part by non-ideal conditions).  We used
 the GBT Ka--band receiver and the Caltech Continuum Backend (CCB).
 For a detailed description of this configuration of equipment and the
 observing methods used, see \cite{Mason2009}; we will describe the
 salient points.

\subsubsection{Calibration}

Every observing session included at least one flux calibration
observation, using the most accessible of the GBT standard calibrators
3C~147, 3C~48 or 3C~286.  We observed sources cluster by cluster, and
thus they tend to lie in small regions.  The GBT pointing is stable
for up to 40\,minutes for night time observing, but we rarely required
re-pointing near a field because slews tended to occur after only a
few source integrations.  We chose one pointing calibrator per cluster
(and hence its associated sources).  We took pointing calibrators from
the GBT database, choosing the brightest available within 10\,degrees
of the cluster centre.  We also performed at least one sky dip per
observing session to establish the sky temperature and opacity.

\setlength{\dashlinedash}{0.5pt}
\setlength{\dashlinegap}{6pt}
\setlength{\LTcapwidth}{7in}

\setlength{\extrarowheight}{3pt}

\begin{table*}
\caption{Radio sources found within 5 arcmin of the cluster pointing
centre in the NVSS \protect\citep{NVSS} and GB6 \protect\citep{GB6}
surveys, and also in the observations of
\protect\cite{Coble2007}. Notes: All sources were observed with the
GBT unless marked (*).  30\,GHz flux densities were derived from our
GBT measurements plus supplementary information at other frequencies
as indicated in the final column (OB for OVRO/BIMA, N for NVSS, G for
GB6 and F for FIRST).  Quoted errors are $\pm 1 \sigma$ ($68\%$
confidence interval).  Some sources were observed more than once with
OVRO/BIMA.  Sources showing evidence for variability are marked (V).
Sources with apparently negative 30\,GHz flux densities are shown in
italics: these data were not used to correct the SZ measurements.}

\begin{center}
\label{table:sources}
\begin{tabular}{llccr@{}lr@{}lr@{}lr@{}ll}
\hline
Cluster &Src &RA &Dec &\multicolumn{2}{c}{$S_{1.4}$}
&\multicolumn{2}{c}{$S_{4.85}$} &\multicolumn{2}{c}{$S_{28.5}
$} &\multicolumn{2}{c}{$S_{30}$} &Notes \\
&\# &(J2000) &(J2000) &\multicolumn{2}{c}{(mJy)} 
&\multicolumn{2}{c}{(mJy)} &\multicolumn{2}{c}{(mJy)} 
&\multicolumn{2}{c}{(mJy)} \\
\hline

A1835 &1* &14 01 02.1 &02 52 41.0 &39.3{}&$\pm1.6$ &-&
&3.31{}&$\pm0.14$ &2.93{}&$\pm0.18$ &OB N\\
& & & & & & & &2.88 {}&$\pm0.07$ & \\
\hdashline
&2* &14 01 00.5 &02 51 53.0 &1.6{}&$\pm0.1$ &-& 
&1.26{}&$\pm0.14$ &1.33{}&$\pm0.07$ &OB F\\
& & & & & & & &1.36{}&$\pm0.08$ & \\
\hline
ZWCL1953 &{1} &{08 50 07.8} &{36 04 21.7}
&{19.4}{}&${\pm1.2}$ &-& &-& &{0.79}{}&{$\pm0.70$}
&{GBT}\\ \hdashline
&2 &08 50 13.4 &36 04 22.4 &20.8{}&$\pm1.0$  &19.0{}&$\pm4.0$
&1.19&{}$\pm0.17$ &1.28{}&$\pm0.16$ &GBT OB\\
\hline
ZW3146 &1 &10 23 37.2 &04 09 06.4 &31.5{}&$\pm1.0$ &-&
&2.03{}&$\pm0.22$ &2.35{}&$\pm0.81$ &GBT OB\\
& & & & & & & &2.12{}&$\pm0.15$ & \\
\hdashline
&2 &10 23 39.6 &04 11 15.4 &7.1{}&$\pm0.5$ &-&
&0.41{}&$\pm0.07$ &0.33{}&$\pm0.07$ &GBT OB\\
\hdashline
&3 &10 23 45.1 &04 10 40.7 &95.8{}&$\pm3.4$ &42.0{}&$\pm7.0$
&5.35{}&$\pm0.18$ &5.27{}&$\pm0.11$ &GBT OB\\
& & & & & & & &5.70{}&$\pm0.10$ & \\
\hdashline
&4 &10 23 45.2 &04 11 39.7 &3.6{}&$\pm0.4$ &-&
&0.85{}&$\pm0.10$ &0.80{}&$\pm0.10$ &GBT OB\\
\hline
RXJ1532.9+0005 &1 &15 32 47.4 &30 18 46.0 &18.0{}&$\pm1.0$ &-&
&-& &1.18{}&$\pm0.21$ &GBT\\
\hdashline
&2 &15 32 50.7 &30 19 47.6 &7.9{}&$\pm0.5$ &-&
&6.58{}&$\pm0.20$ &6.56{}&$\pm0.21$ &OB N\\
\hdashline
&3 &15 32 53.8 &30 20 59.8 &22.8{}&$\pm0.8$ &20.0{}&$\pm4.0$
&3.25{}&$\pm0.18$ &3.27{}&$\pm0.12$ &GBT OB\\
\hdashline
&4 &15 32 54.31 &30 23 01.5 &4.4{}&$\pm0.4$ &-&
&-& &0.92{}&$\pm0.21$ &GBT\\
\hdashline
&5 &15 33 03.42 &30 23 47.2 &2.6{}&$\pm0.4$ &-&
&-& &1.12{}&$\pm0.50$ &GBT\\
\hline
A2390 &1 &21 53 32.4 &17 42 19.8 &20.9{}&$\pm1.6$ &-&
&\multicolumn{2}{c}{Not observed} &0.23{}&$\pm0.12$ &GBT\\
\hdashline
&2 &21 53 36.8 &17 41 44.8 &235.3{}&$\pm8.3$ &220{}&$\pm20$
&\multicolumn{2}{c}{Not observed} &45.77{}&$\pm0.10$ &GBT\\
\hdashline
&\emph{3} &\emph{21 53 40.3} &\emph{17 42 56.7} &\emph {12.2}{}&$\it \pm1.0$
&-& &\multicolumn{2}{c}{\emph{Not observed}} &-\emph{0.14}{}&$\it \pm0.09$ 
&\emph{GBT}\\
\hline
A2219 &1 &16 40 21.83 &46 42 47.8 &239.1{}&$\pm8.3$ &84.8{}&$\pm8.0$
&14.87{}&$\pm0.17$ &13.74{}&$\pm1.00$ &GBT OB\\
\hdashline
&2* &16 40 23.83 &46 41 47.3 &7.9{}&$\pm1.0$ &-& &0.97{}&$\pm0.17$
&0.94{}&$\pm0.17$ &OB N\\
\hdashline
&3 &16 40 14.98 &46 42 28.7 &6.1{}&$\pm0.5$ &-& &-& &0.42{}&$\pm0.09$
&GBT\\ 
\hdashline
&4 &16 39 58.07 &46 40 37.2 &14.1{}&$\pm0.5$ &-& &-& &0.96{}&$\pm0.09$
&GBT\\
\hline
RXJ2129.6+0005 &\emph{1} &\emph{21 29 36.61}  &\emph{00 02 35.4}
&\emph{2.2}{}&$\it \pm0.4$ &-& &-& &-\emph{0.13}{}&$\it \pm0.21$ &\emph{GBT}\\
\hdashline
&2 &21 29 40.00  &00 05 22.9 &25.4{}&$\pm1.2$ &-& &2.33{}&$\pm0.10$
&2.04{}&$\pm0.06$ &GBT OB\\
\hdashline
&{3} &{21 29 40.14}  &{00 01 44.4} &{4.5}{}&$ \pm0.5$ 
&-& &-& &{0.11}{}&$ \pm0.13$ &{GBT}\\
\hdashline
&4 &21 29 55.24  &00 07 56.9 &34.3{}&$\pm1.8$ &-& &-& 
&2.82{}&$\pm0.11$ &GBT\\
\hline
A2261 &1 &17 22 16.9 &32 09 10.5 &23.0{}&$\pm1.5$ &-& &9.32{}&$\pm0.22$
&15.58{}&$\pm3.89$ &GBT OB\,\,(V)\\
&&&&&&& &10.48{}&$\pm0.16$ & &\\
\hdashline
&2 &17 22 27.7 &32 07 57.8 &5.3{}&$\pm0.5$ &-& &-& &0.13{}&$\pm0.08$ &GBT\\
\hdashline
&3 &17 22 45.3 &32 09 27.0 &4.85{}&$\pm0.5$ &-& &-& &0.14{}&$\pm0.10$ &GBT\\
\hdashline
&4 &17 22 06.55 &32 07 01.7 &5.8{}&$\pm0.4$ &-& &-& &0.14{}&$\pm0.08$ &GBT\\
\hline
A781 &1 &09 20 08.51 &30 32 14.3 &19.9{}&$\pm0.9$ &-& &-& &1.42{}&$\pm0.60$
&GBT\\
\hdashline
&2 &09 20 14.12 &30 29 02.8  &17.9{}&$\pm0.7$ &-& &-& &1.22{}&$\pm0.53$
&GBT\\
\hdashline
&3 &09 20 21.75 &30 32 27.0 &2.7{}&$\pm0.5$ &-& &-& &0.11{}&$\pm0.18$ &GBT\\
\hdashline
&4 &09 20 22.90 &30 29 45.6 &73.1{}&$\pm2.6$ &32.0{}&$\pm5.0$ 
&5.33{}&$\pm0.18$ &5.35{}&$\pm0.30$ &GBT OB\\
\hdashline
&\emph{5} &\emph{09 20 30.83} &\emph{30 28 02.1} &\emph{15.8}{}&$\it\pm1.6$
&-& &-& &\emph{-0.07}{}&$\it\pm0.16$ &\emph{GBT}\\
\hdashline
&6 &09 20 47.35 &30 28 21.3 &5.4{}&$\pm0.6$ &-& &-& &0.08{}&$\pm0.17$ &GBT\\
\hline
A697 &1 &08 42 40.22 &36 19 16.4 &5.4{}&$\pm0.5$ &-& &-& &1.64{}&$\pm0.54$
&GBT\\
\hdashline
&2 &08 42 59.67 &36 17 43.7 &32.5{}&$\pm1.4$ &-& &-& &1.71{}&$\pm0.65$ &GBT\\
\hline
\end{tabular}
\end{center}
\end{table*}
\begin{table*}
\contcaption{}
\begin{center}
\begin{tabular}{llccr@{}lr@{}lr@{}lr@{}ll}
\hline
Cluster &Src &RA &DEC &\multicolumn{2}{c}{$S_{1.4}$}
&\multicolumn{2}{c}{$S_{4.85}$} &\multicolumn{2}{c}{$S_{28.5}
$} &\multicolumn{2}{c}{$S_{30}$} &Notes \\
&\# &(J2000) &(J2000) &\multicolumn{2}{c}{(mJy)} 
&\multicolumn{2}{c}{(mJy)} &\multicolumn{2}{c}{(mJy)} 
&\multicolumn{2}{c}{(mJy)} \\
& & & &\multicolumn{2}{c}{(NVSS)}
&\multicolumn{2}{c}{(GB6)} &\multicolumn{2}{c}{(O/B)} 
&\multicolumn{2}{c}{(GBT+)}\\
\hline
A1763 &\emph{1} &\emph{13 35 15.60} &\emph{41 00 25.8} 
&\emph{7.9}{}&$\it\pm0.5$ &-& &-& &\emph{-0.10}{}&$\it\pm0.14$
&\emph{GBT}\\
\hdashline
&2* &13 35 19.86 &41 00 04.0 &859.2{}&$\pm29.5$ &225{}&$\pm20$
&31.30{}&$\pm0.41$ &29.49{}&$\pm0.38$ &OB G N\\
\hline
A68 &1 &00 36 52.94 &09 05 21.7 &22.8{}&$\pm0.8$ &-& &-& &1.57{}&$\pm0.05$
&GBT\\
\hdashline
&2 &00 37 05.21 &09 13 33.6 &4.4{}&$\pm0.4$ &-& &-& &0.60{}&$\pm0.05$ &GBT\\  
\hdashline
&3 &00 37 06.35 &09 07 30.4 &40.2{}&$\pm1.3$ &-& &1.20{}&$\pm0.12$
&1.21{}&$\pm0.06$ &GBT OB\\
\hdashline
&4 &00 37 07.71 &09 08 24.0 &59.1{}&$\pm2.2$ &-& &1.60{}&$\pm0.10$
&1.65{}&$\pm0.05$ &GBT OB\\
\hdashline
&\emph{5} &\emph{00 37 17.93} &\emph{09 06 30.5} &\emph{3.2}{}&$\it\pm0.6$ 
&-& &-& &\emph{-0.05}{}&$\it\pm0.04$ &\emph{GBT} \\
\hdashline
&6* &00 37 07.00 &09 07 58.7 &-& &-& &1.38{}&$\pm0.11$ &-& &OB\\
\hline
A520 
&1 &04 54 01.11 &02 57 45.6 &6.3{}&$\pm0.5$  &-& &7.83{}&$\pm0.25$ 
&5.75{}&$\pm1.74$ &GBT OB (V)\\ 
&&&&&&& &4.42{}&$\pm0.88$ & &\\
\hdashline
&2 &04 54 03.72 &02 56 01.9 &7.7{}&$\pm1.7$  &-& &-& &0.49{}&$\pm0.15$ &GBT\\
\hdashline
&\emph{3} &\emph{04 54 12.41} &\emph{02 57 52.3} &\emph{7.1}{}&$\it\pm0.5$  
&-& &-& &\emph{-0.15}{}&$\it\pm0.16$
&\emph{GBT}\\
\hdashline
&4 &04 54 17.14 &02 55 33.5 &14.1{}&$\pm1.0$ &-& &0.84{}&$\pm0.11$ 
&0.90{}&$\pm0.08$ &GBT OB\\
&&&&&&& &1.09{}&$\pm0.09$ \\
\hdashline
&5 &04 54 21.16 &02 55 01.4 &26.5{}&$\pm1.6$ &-& &1.00{}&$\pm0.13$
&0.63{}&$\pm0.10$ &GBT OB\\
&&&&&&& &0.74{}&$\pm0.14$\\
\hline
A267 &1 &01 52 29.46 &00 59 31.8 &30.0{}&$\pm1.0$ &-& &2.75{}&$\pm0.20$ 
&3.12{}&$\pm0.23$ &GBT OB\\
\hdashline
&2 &01 52 54.58 &01 02 08.2 &4.2{}&$\pm0.5$ &-& &7.55{}&$\pm0.24$ 
&7.05{}&$\pm1.12$ &GBT OB(V)\\
&&&&&&& &5.53{}&$\pm0.73$ \\
\hdashline
&3 &01 52 34.35 &01 01 20.5 &-& &-& &-& &0.59{}&$\pm0.32$ &GBT\\
\hline
RXJ0439.0+0715 &1 &04 39 01.26 &07 15 42.6 &30.6{}&$\pm1.4$ &-& 
&1.18{}&$\pm0.16$ &0.89{}&$\pm0.22$ &GBT OB\\
\hline
ZWCL7160 
&1 &14 57 08.06 &22 20 11.2 &13.2{}&$\pm0.6$ &-& &0.95{}&$\pm0.05$
&0.90{}&$\pm0.06$ &GBT OB\\
\hdashline
&2 &14 57 08.06 &22 20 11.2 &3.9{}&$\pm0.4$ &-& &0.99{}&$\pm0.07$ 
&0.91{}&$\pm0.08$ &GBT OB\\
\hdashline
&3 &14 57 15.18 &22 20 36.0 &16.5{}&$\pm1.3$ &-& &0.96{}&$\pm0.04$
&0.90{}&$\pm0.06$ &GBT OB\\
\hline
A773 &1* &09 17 45.39 &51 43 11.2 &2.7{}&$\pm0.5$ &-& &\multicolumn{2}{c}{$<0.39$} &-& 
&\emph{Negligible}\\
\hdashline
&2* &09 18 01.81 &51 44 11.4 &3.1{}&$\pm0.5$ &-& &\multicolumn{2}{c}{$<0.39$} &-&
&\emph{Negligible}\\
\hline
\\

\setlength{\extrarowheight}{0pt}

\end{tabular}
\end{center}
\end{table*}

\subsubsection{Differencing}

We employed the standard GBT nodding strategy, which is similar to the
OCRA differencing scheme, in order to remove atmospheric
contamination.  A complete nod (ie the sequence of beam 1 on source,
beam 2 on source, beam 2 on source, beam 1 on source) takes
1.5\,minutes.  Each source was observed for at least one complete
nod.  The number of nods was governed by telescope scheduling,
though we tried to include extra nods for sources expected to have
lower flux densities.

\subsubsection{GBT flux density determinations}

The CCB has a 14\,GHz bandwidth, split into four frequency sub--bands
centred at 27.75, 31.25, 34.75 and 38.25\,GHz. We can thus measure
simultaneously four separate flux densities for each source, from
which we can derive a 30\,GHz value by fitting a power law to the
available data. We use a Monte--Carlo method to estimate the
uncertainties on the interpolated flux densities.  For poor--quality
data, we were unable to derive accurate 30\,GHz flux densities from
the GBT data alone, as discussed in Section \ref{sec:add}

\subsection{Determining the 30\,GHz flux densities}
\label{sec:add}
The quality of our GBT data is rather mixed due to the wide range of
observing conditions experienced. As most sources were only observed
once, we are also unable to constrain source variability with GBT data
alone.  In order to supplement our 30\,GHz measurements, we make use
of the flux densities reported in \cite{Coble2007} who observed all of
our clusters except Abell 2390 at 28.5\,GHz with the OVRO and BIMA
interferometers.  Such interferometric data not only measure the SZ
effect but also enable identification of point sources via the longest
baselines, where the SZ signal will be negligible.  The uncertainties
on the \citeauthor{Coble2007} flux densities are generally a few
tenths of a mJy, so given the proximity of the two observing
frequencies, we are able to place tighter constraints on our 30\,GHz
source flux densities by fitting to both datasets, which is
particularly useful for sources with poor GBT data.  In addition, we
are able to identify variable sources and estimate the effect on our
SZ measurements.

Most source flux densities are well constrained by a combination of GBT
and OVRO/BIMA data.  We fit a power law to the measurements in the four
GBT frequency sub-bands, plus the interferometer data at 28.5\,GHz.  For
sources with good GBT data, the \citeauthor{Coble2007} points serve as a
valuable check for variability, but otherwise have minimal effect. In
the three cases where significant variability is found, we take the two
differing flux density measurements, and use half the difference as an
estimate of the uncertainty in our measurement due to variability. We
then add this in quadrature to the measurement error.  The source
exhibiting the greatest variability, source~1 in Abell~2261, has little
effect on our results since it lies well away from the cluster centre
and data where it lies close to a reference arc are flagged out (Section
\ref{sec:corr}).  Where no 28.5\,GHz flux density is available, we
proceed with the GBT data alone.  Such cases are unlikely to be crucial,
as the non detection by OVRO/BIMA suggests the source is either rather
faint, or well away from the central regions of the cluster.  Many
sources have GBT data which are usable but rather noisy.  In such cases,
the 28.5\,GHz data are invaluable in tying down the 30\,GHz flux density
due to their comparatively small uncertainties.

Seven sources have no usable GBT data: two in A1835, one in A2219, one
in A1763, one in A68 and two in A773.  We now discuss our method for
each source in turn.  Source 1 in A1835 was measured twice by
OVRO/BIMA, and also appears in NVSS.  We fit a power law to the three
measurements.  As our observing frequency is close to that of
OVRO/BIMA, assuming a constant spectral index is unlikely to have a
large impact.  We rule out significant variability.  Source 2 in A1835
was also measured twice at 28.5\,GHz, but does not appear in NVSS.  We
consult FIRST for a lower frequency measurement, and proceed as for
source 1.  For source 2 in A2219, we again fit a power law to the
OVRO/BIMA and NVSS datapoints.  Source 6 in A68 does not appear in any
of the lower frequency catalogues but is well measured by
OVRO/BIMA. It is outside the OCRA beams so its flux density has only a
small effect on the SZ measurement.  Both sources in A773 are faint in
NVSS and are not detected by OVRO/BIMA.  The noise on the OVRO map is
$0.078$\,mJy, giving a $5 \sigma$ upper limit of $0.39$\,mJy.  As the
sources in A773 lie well away from the OCRA beams, we are confident
that their effects on our data are negligible.

\subsection{Source corrections}
\label{sec:corr}
 
Due to OCRA-p's position switching strategy, contaminant radio sources
can affect the data by producing a \emph{positive} signal when they
lie close to the cluster centre, or a rogue \emph{negative} signal
when they lie in the reference beam of the telescope. We simulate the
effects of the sources based on their measured positions and flux
densities, and the OCRA beam response.  We then use the resulting
effective flux density to correct the SZ data.  Contamination through
sources at 30\,GHz is generally low for bright clusters, producing
only small corrections to the SZ data, and thus introduces little bias
as a result of the limitations of the method employed.  Although we
are confident that residual biases are below the level of our errors,
we recognise that the absence of 30\,GHz data is a limitation in some
cases, and there may be additional issues to do with source
variability.  For clusters where radio sources fall in the OCRA
reference beams, we are able to perform an additional check by
flagging out contaminated data based on the range of parallactic
angles affected by the source in question.  We compare the flagged
data with the corrected data, and adjust our final measurement error
accordingly. 

We recognise that our strategy may miss sources which are important at
30\,GHz.  The flux limit of NVSS is $\sim2.5$\,mJy.  If we consider a
source of flux density $\sim2.4$\,mJy at 1.4\,GHz, and assume a
typical spectral index of $-0.7$, extrapolating to 30\,GHz gives a
flux density of $0.34$\,mJy.  Even if such a source were to be located
in the central regions of a cluster, the effect on our measurement
would still be small relative to random errors.  Not all sources have
typical spectral indices, but we note that in the 16 clusters
additionally observed by \cite{Coble2007}, only 1 `new' source, i.e. a
source not previously identified in NVSS, was found.  However, the
quality of our knowledge of the population of sources with mJy flux
densities at 30\,GHz is poor.  We were able to run an additional check
for `new' sources lying in OCRA--p's reference arcs. We binned the
data by parallactic angle to look for contaminating sources that might
produce false SZ effects. Although this subdivides the data, and so
leads to a noise level that is higher by a factor of $2 - 3$ than the
central measured flux density averaged over all parallactic angles, no
previously uncatalogued sources were detected.


\section{X-RAY DATA}

\subsection{Deriving cluster parameters}
\label{sec:X}

Our sample has 15 clusters in common with an upcoming X-ray paper,
Giles et al. (in prep), some results from which were made available in
advance for this work.  The remaining two clusters (ZWCL1953 and
RXJ1532.9+3021, which lie outside Giles et al.'s specified redshift
range of $0.15<z<0.30$) were analysed using an identical method.  The
authors use \emph{Chandra} ACIS--I and ACIS--S imaging and
spectroscopy for each cluster as available via the archive.  For the
data preparation, they follow closely the method outlined in
\citet{Maughan2008}. The \emph{Chandra} analysis package used was
CIAO\footnote{http://cxc.harvard.edu/ciao/} version 4.2 and
calibration database CALDB 4.1.4.  As in \citet{Maughan2008}, blank
sky, rather than local, backgrounds are used because the clusters fill
a large fraction of the field of view.

We fit a gas density model by converting the observed surface
brightness profile into a projected emissivity profile, which was then
modelled by projecting the modifed $\beta$--model of
\citealp{Vikhlinin2006} along the line of sight (see
\citealp{Maughan2008} for details).  Gas parameters and errors were
determined from Monte Carlo realisations of the projected emissivity
profile.  At each data point a new randomised point was drawn from a
Gaussian distribution centered on the model value at that point, with
a standard deviation equal to the fractional measurement error on the
original data point, multiplied by the model value. The error bar on
this new randomised point was set to the same fractional size as the
error bar on the original observed point.

The cluster temperature, gas mass and $R_{500}$ (the radius enclosing
a mean density of 500 times the critical density at the cluster's
redshift) were then determined iteratively. The procedure followed was
to extract a spectrum from within an estimated $R_{500}$ (with the
central $15\%$ of that radius excluded), integrate the gas density
profile to determine the gas mass within the estimated $R_{500}$, and
thus calculate $Y_\mathrm{X}$ which is the product of $kT$ and the gas
mass, and a low scatter proxy for total mass \citep{Kravtsov2006}. A new
value of $R_{500}$ was then estimated from the $Y_\mathrm{X}$-$M$
scaling relation of \citet{Vikhlinin2009},

\begin{eqnarray}\label{e.ym}
M_{500} & = & E(z)^{-2/5}\,A_\mathrm{YM}\left(\frac{Y_X}{3\times10^{14}
M_{\odot}\mathrm{keV}}\right)^{B_\mathrm{YM}},
\end{eqnarray}

\noindent with $A_\mathrm{YM}=5.77\times10^{14}h^{1/2}M_{\odot}$ and
$B_\mathrm{YM}=0.57$. Here, $M_{500}$ is the mass within $R_{500}$ (allowing
$R_{500}$ to be trivially computed), and

\begin{equation}
E(z)=\sqrt{\Omega_{\mathrm{M}}(1+z)^3 +
(1-\Omega_{\mathrm{M}}-\Omega_\Lambda)(1+z)^2 + \Omega_\Lambda}
\label{eqn:E}
\end{equation}

\noindent describes the redshift evolution of the Hubble parameter.
This $Y_{\mathrm{X}}$--$M$ relation assumes self-similar evolution (as
$E^{-2/5}$), which has been shown to be a good description of observed
clusters to $z\approx0.6$ \citep{Maughan2007}. The process was repeated
until $R_{500}$ converged.  In these spectral fits, the source emission
was modelled with an absorbed thermal plasma APEC model in the
0.6--9\,keV band, with the absorbing column fixed at the Galactic value.
$N_\mathrm{H}$ values were taken from the HI map by
\citet{Kalberla2005}.

Our final cluster parameters, with $T_\mathrm{X}$, $L_\mathrm{X}$ and
$Y_\mathrm{X}$ determined within a radius of $R_{500}$ are given
in Table \ref{tab:X}.

\begin{table*}
\caption{\emph{Chandra} X-ray parameters for the sample, derived
inside $R_{500}$ with the central 15\% excluded. $L_{\mathrm{X}}$ is
the bolometric luminosity. Quoted errors are $\pm 1 \sigma$ ($68\%$
confidence interval).}
\begin{tabular}{lr@{}lr@{}lr@{}lc}
\hline 
Cluster &\multicolumn{2}{c}{$T_{\mathrm{X}}$}
&\multicolumn{2}{c}{$L_{\mathrm{X}}$} 
&\multicolumn{2}{c}{$Y_{\mathrm{X}}$} &$R_{500}$\\ 
&\multicolumn{2}{c}{(keV)}
&\multicolumn{2}{c}{($10^{45}\mathrm{erg}\mathrm{s}^{-1}$)}
&\multicolumn{2}{c}{($10^{15}\mathrm{M_{\odot}keV}$)} &Mpc\\
\hline
A1835 &8.41&${}^{+0.32}_{-0.26}$ &2.14&{}$\pm0.01$ 
&1.20&${}^{+0.05}_{-0.04}$ &1.40\\
ZW1953 &6.05&{}$^{+0.51}_{-0.50}$ &1.45&{}$\pm0.05$
&0.58&${}^{+0.05}_{-0.05}$ &1.19\\
ZW3146 &7.09&{}$^{+0.35}_{-0.37}$ &1.62&{}$\pm0.02$
&0.77&{}$^{+0.04}_{-0.04}$ &1.27\\ 
RXJ1532.9+3021 &5.62&{}$^{+0.72}_{-0.58}$ &1.23&{}$\pm{0.05}$
&0.45&{}$^{+0.06}_{-0.06}$ &1.11\\
A2390 &10.20&{}$^{+0.35}_{-0.35}$ &2.87&{}$\pm0.05$
&1.79&{}$^{+0.06}_{-0.07}$ &1.53\\
A2219 &10.33&{}$^{+0.43}_{-0.44}$ &3.32&{}$\pm0.04$
&2.10&$^{+0.09}_{-0.09}$ &1.58\\ 
RXJ2129.6+0005 &6.04&{}$^{+0.53}_{-0.52}$ &1.03&{}$\pm0.04$
&0.52&{}$^{+0.05}_{-0.06}$ &1.20\\
A2261 &6.56&{}$^{+0.28}_{-0.28}$ &1.46&{}$\pm0.02$
&0.77&{}$^{+0.03}_{-0.03}$ &1.31\\
A781 &5.32&{}$^{+0.66}_{-0.43}$ &1.02&{}$\pm0.05$ 
&0.53&{}$^{+0.06}_{-0.04}$ &1.18\\
A697 &9.22&{}$^{+0.66}_{-0.65}$ &2.70&{}$\pm0.05$
&1.55&{}$^{+0.11}_{-0.11}$ &1.45\\
A1763 &7.31&{}$^{+0.47}_{-0.47}$ &1.48&$\pm0.03$
&0.89&{}$^{+0.06}_{-0.06}$ &1.34\\
A68 &7.14&{}$^{+0.88}_{-0.78}$ &1.09&{}$\pm0.04$
&0.60&{}$^{+0.07}_{-0.07}$ &1.23\\
A520 &6.78&{}$^{+0.29}_{-0.16}$ &1.49&{}$\pm0.02$              
&0.74&{}$^{+0.03}_{-0.02}$ &1.31\\
A267 &4.53&{}$^{+0.49}_{-0.38}$ &0.75&{}$\pm0.05$	       
&0.29&{}$^{+0.03}_{-0.03}$ &1.08\\
RXJ0439.0+0715 &5.32&$^{+0.38}_{-0.29}$ &1.58&$\pm0.04$              
&0.57&{}$^{+0.04}_{-0.04}$ &1.23\\
ZW7160 &4.64&{}$^{+0.18}_{-0.18}$ &0.69&{}$\pm0.01$	       
&0.28&{}$^{+0.01}_{-0.01}$ &1.06\\	 
A773 &6.75&{}$^{+0.40}_{-0.26}$ &1.13&$\pm0.02$              
&0.65&{}$^{+0.03}_{-0.03}$ &1.27\\
\hline
\end{tabular}
\label{tab:X}
\end{table*}

\subsection{$\beta$--model fitting}
\label{sec:beta}

As mentioned in Section \ref{sec:obs}, the small separation of the
OCRA-p beams relative to the extent of a cluster's SZ decrement
implies that we subtract out a fraction of the SZ signal due to our
differencing method.  This significantly reduces the measured signal
compared to the intrinsic central signal.  This is further exaggerated
in cases where we have adjusted the pointing centre to avoid bright
radio sources.  In order to remove this effect and recover the `true'
SZ signal, we model the cluster atmosphere and estimate the fraction
of the signal `missed'.  (Note: this analysis was conducted purely to
correct the SZ measurements and is separate from the derivation of
X--ray cluster parameters discussed in Section \ref{sec:X}).  In our
previous work (\citealp{Lancaster2007}), we assumed that the gas
density distribution, $n_{\mathrm{e}}(r)$ was well fit by a standard
spherical $\beta$--model \citep{beta1}:

\begin{equation}
n_{\mathrm{e}}(r) = n_{\mathrm{e0}}\left(1+ \frac{r^2}{r^2_c}
\right)^{-3\beta/2}
\label{eq:single}
\end{equation}

\noindent Here, we improve on this approach by also considering the
double $\beta$--model, which may provide a better fit to clusters
which exhibit peaked central emission due to their cool cores.  We
follow \citet{Bonamente2006} in adopting the form

\begin{equation}
n_{\mathrm{e}}(r) = n_{\mathrm{e0}} \left[p\left(1+
\frac{r^2}{r^2_{c1}}\right)^{-3\beta_1/2} +(1-p)\left(1+
\frac{r^2}{r^2_{c2}}\right)^{-3\beta_2/2} \right]
\label{eq:double}
\end{equation}

\noindent although in our analysis, we allow different $\beta$ for the
two components.  Using the
\emph{SHERPA}\footnote{http://cxc.harvard.edu/sherpa/} modelling and
fitting application for each set of cluster data, we fit the 1--D
radial profiles of the X--ray surface brightness by three models:
\begin{enumerate}
\item{The standard single $\beta$--model}, (\ref{eq:single})
\item{A double $\beta$--model, (\ref{eq:double}), with $\beta_1 = \beta_2$}
\item{A double $\beta$--model, (\ref{eq:double}), allowing $\beta_1$ and $\beta_2$ to
assume different values.}
\end{enumerate}

\noindent Chi squared ($\chi ^2$) statistics are used to compare the
fits, since each radial bin is based on a large number of X--ray
counts.  For each cluster, we accept the simplest model for which the
fit statistic implies a probability of getting a greater $\chi ^2$ by
chance, $Q \ge 6$ per cent.  For some clusters, the gas distribution
is sufficiently disturbed that $Q < 6 $ per cent for all models
fitted: in such cases we adopt the best fit.  This is not ideal, but
we expect the additional error to be at a low level in comparison with
the random errors on the SZ data.  We do not observe any systematic
effect on our results due to the type of $\beta$--model used to
correct the SZ data.

$\beta$--model parameters are presented in Table \ref{tab:beta}.  We
return briefly to the analysis of \cite{Bonamente2006}, with whom we
have eight clusters in common, for comparison of results.  For A697,
A68, A267 and A773, both groups find that a single $\beta$--model
provides an acceptable fit, and the derived parameters are in broad
agreement.  For their double $\beta$--model fits,
\citeauthor{Bonamente2006} fix the two values to $\beta$ to be equal,
whereas ours were allowed to vary.  For RXJ2129.6+0005 and A2261, we
find that a model with $\beta_1=\beta_2$ gives the best fit, and the
model parameters are again in broad agreement with
\citeauthor{Bonamente2006} For ZWCL3146, our two $\beta$--values
differ but are very similar, meaning the overall model fit is in broad
agreement with \citeauthor{Bonamente2006}, and indeed our
$\beta_1=\beta_2$ fit is also consistent. The exception is A1835,
where we find the best fitting model to be a double $\beta$--model
with $\beta_1=0.675$, $\beta_2=1.059$, $\theta_1=35.6$\,arcsec and
$\theta_2=13.2$\,arcsec, whereas \citeauthor{Bonamente2006} find
$\beta=0.797$, $\theta_1=63.6$\,arcsec and $\theta_2=13.2$\,arcsec.  Our
model with $\beta_1=\beta_2$ is also slightly discrepant, as our best
fit parameters are $\beta=0.695$, $\theta_1=92.1$\,arcsec and
$\theta_2=19.1$\,arcsec.  The source of discrepancy may well be in the
use of differing data sets.  We use ACIS-I observations totalling
200ks, whereas \citeauthor{Bonamente2006} only had access to 30ks of
ACIS-S data which has a significantly smaller field of view.

We use the model gas distribution to simulate an OCRA observation.
Many clusters were observed at positions slightly shifted from their
X-ray maxima in order to avoid bright radio sources as far as
possible.  We consult our OCRA data for the range of parallactic
angles sampled, and include the convolution with the OCRA beam
response.  We can thus estimate the fraction $F$ of the `true' SZ
signal measured in the real observations, and `correct' our
measurement (multiply by $1/F$) to recover the SZ surface brightness
that would be measured by an ideal instrument with narrow beams at
infinite separation. We note for clusters where the double
$\beta$--model provides an improved fit, the value of F varies from
the simpler single $\beta$--model by $\sim10$ per cent.  This change
is thus significant for the scientific interpretation of our results.
Correction factors are presented in Table \ref{tab:res}.  Asymmetries
in the SZ signal are assumed small in our analysis.  Checks for such
asymmetries by binning the SZ data by parallactic angle provide only a
low signal--to--noise control against such effects which would cause
our estimates of $F$ to be incorrect.  The X--ray isophotes of these
clusters are not always circular, but are most sensitive to asymmetry
towards the cluster centre, where a relatively small fraction of the
SZ effect originates.  We estimate that the value of $F$ could be
systematically uncertain by $20\%$ from this effect.  This uncertainty
dominates over that from varying the density model from $\beta$ or
double-$\beta$ form to the \cite{Vikhlinin2006} form.

\begin{table*}
\caption{$\beta$--model parameters for the cluster sample.  The best
fitting model is shown for each cluster, where the parameter $p$
indicates the fraction of each component as described by Equation
\ref{eq:double}. Quoted errors are $\pm 1 \sigma$ ($68\%$ confidence
interval).  In cases where a double--$\beta$ model provided the best
fit, we choose the larger of the core radii to be $r_{c1}$ The
resulting correction factors applied to the SZ data are given in Table
\ref{tab:res}.}
\begin{tabular}{lr@{}lr@{}lr@{}lr@{}lc}
\hline
Cluster &\multicolumn{2}{c}{$\beta_1$} &\multicolumn{2}{c}{$r_{c1}$} 
&\multicolumn{2}{c}{$\beta_2$} &\multicolumn{2}{c}{$r_{c2}$} &$p$\\

& &&\multicolumn{2}{c}{(arcsec)} 
& &&\multicolumn{2}{c}{(arcsec)} \\
\hline
A1835 &0.675{}&$^{+0.001}_{-0.001}$ &35.6{}&$^{+0.6}_{-0.6}$
&1.059&{}$^{+0.010}_{-0.010}$ &13.2{}&$^{+0.1}_{-0.1}$ &0.10 \\
ZWCL1953 &0.972{}&$^{+0.147}_{-0.122}$ &171.6&{}$^{+22.3}_{-21.9}$
&0.972{}&$^{+0.147}_{-0.122}$ &43.3&{}$^{+5.7}_{-5.1}$ &0.04 \\
ZWCL3146 &0.700&{}$^{+0.008}_{-0.007}$ &23.6&{}$^{+0.9}_{-0.8}$
&0.652&{}$^{+0.032}_{-0.023}$ &4.8&{}$^{+0.3}_{-0.2}$ &0.14 \\
RXJ1532.9+3021 &0.650&{}$^{+0.011}_{-0.010}$ &12.0&{}$^{+3.6}_{-1.7}$
&0.650&{}$^{+0.011}_{-0.010}$ &6.3&{}$^{+1.6}_{-1.8}$ &0.40 \\
A2390 &0.831&{}$^{+0.010}_{-0.009}$ &92.9&{}$^{+1.6}_{-1.6}$
&0.831&{}$^{+0.010}_{-0.009}$ &18.5&{}$^{+0.3}_{-0.3}$ &0.10 \\
A2219 &0.725&{}$^{+0.015}_{-0.013}$ &85.6&{}$^{+6.2}_{-5.0}$
&0.725&{}$^{+0.015}_{-0.013}$ &42.0&{}$^{+5.4}_{-5.6}$ &0.58 \\
RXJ2129.6+0005 &0.609&{}$^{+0.011}_{-0.011}$ &24.4&{}$^{+2.9}_{-2.5}$
&0.609&{}$^{+0.011}_{-0.011}$ &5.0&{}$^{+0.8}_{-0.7}$ &0.12 \\
A2261 &0.598&{}$^{+0.003}_{-0.003}$ &30.7&{}$^{+1.4}_{-1.4}$
&0.598&{}$^{+0.003}_{-0.003}$ &11.8&{}$^{+0.8}_{-0.8}$ &0.33 \\
A781 &0.706&{}$^{+0.041}_{-0.035}$ &79.5&{}$^{+7.5}_{-6.6}$
&-& &-& &- \\
A697 &0.648&{}$^{+0.008}_{-0.008}$ &48.7&{}$^{+1.5}_{-1.5}$
&-& &-& &- \\
A1763 &0.586&{}$^{+0.007}_{-0.006}$ &48.0&{}$^{+1.6}_{-1.6}$
&-& &-& &- \\
A68 &0.742&{}$^{+0.003}_{-0.002}$ &54.5&{}$^{+3.5}_{-3.3}$
&-& &-& &- \\
A520 &0.831&{}$^{+0.003}_{-0.003}$ &117.5&{}$^{+0.6}_{-0.6}$
&-& &-& &- \\
A267 &0.633&{}$^{+0.011}_{-0.011}$ &34.4&{}$^{+1.5}_{-1.4}$
&-& &-& &- \\
RXJ0439.0+0715 &0.735&{}$^{+0.022}_{-0.020}$ &52.6&{}$^{+4.3}_{-3.9}$
&0.735&{}$^{+0.022}_{-0.020}$ &13.0&{}$^{+1.6}_{-1.5}$ &0.23 \\
ZWCL7160 &0.850&{}$^{+0.112}_{-0.077}$ &22.2&{}$^{+4.3}_{-3.4}$
&0.580&{}$^{+0.005}_{-0.006}$ &7.2&{}$^{+0.3}_{-0.4}$ &0.09 \\
A773 &0.769&{}$^{+0.041}_{-0.035}$ &129.6&$^{+14.3}_{-14.4}$
&0.769&{}$^{+0.041}_{-0.035}$ &48.2&{}$^{+2.7}_{-2.6}$ &0.11 \\
\hline
\end{tabular}
\label{tab:beta}
\end{table*}


\section{RESULTS AND SCALING RELATIONS}
\label{sec:res}

\subsection{Sunyaev Zel'dovich Results}

We present our measurements for the Sunyaev Zel'dovich effects in our
17-cluster sample in Table \ref{tab:res}.  For each cluster, we first
give the raw OCRA flux density as measured directly by the telescope,
followed by the revised figure once all radio source contamination has
been corrected for.  We then give our estimate of the fraction of
total signal measured due to the instrumental response.  Finally, we
give the fully--corrected temperature decrement expressed in
brightness temperature units, and the corresponding central
Comptonisation parameter, $y_0 = -0.19 \times (\Delta T/K)$.  We
detect 13 clusters at significance $\ge 3 \sigma$.

\begin{figure*}
\includegraphics[width=0.6\linewidth]{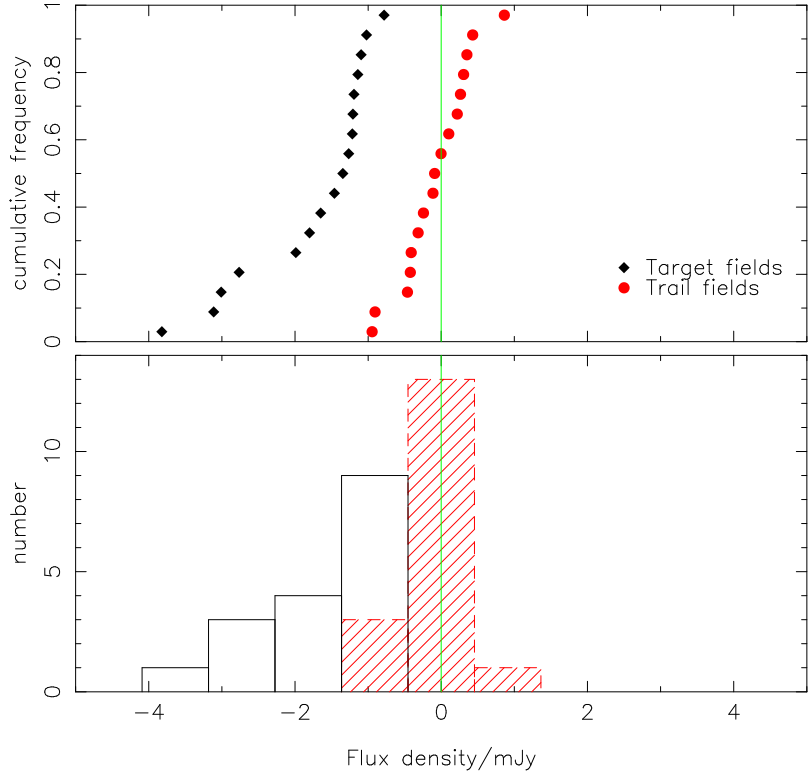}
\caption{Lower panel: histogram of the source-corrected flux
Êdensities for the target cluster fields (black solid lines) and the
Êflux densities for the reference trail fields (red dashed lines with
Êcross-hatching). Upper panel: the corresponding cumulative
Êdistributions for the clusters (black diamonds) and the trail fields
Ê(red circles). Note that the trail field flux densities are consistent
Êwith zero, suggesting that our switching scheme removes atmospheric
Êand ground-based contamination effectively. The cluster and trail
Êfield flux density distributions clearly differ.}
\label{fig:hist}
\end{figure*}

Our trail fields serve as a valuable test of our observing strategy.
We present differential and cumulative distributions for both the source corrected cluster
target and trail fields in Figure \ref{fig:hist}.  The flux densities
of the trail fields are distributed around zero with small scatter,
and their distribution is clearly distinct from that of the SZ data,
confirming that our observing strategy is effective.

\begin{table*}
\caption{OCRA SZ measurements. $S_{\mathrm{Raw}}$ gives the raw OCRA
measurement, whereas $S_{\mathrm{SC}}$ is the SZ flux after correction
for contaminant point sources.  The fourth column lists the measured
fractions of the expected SZ signal due to the OCRA switching
pattern. $\Delta T$ is the brightness temperature decrement, corrected
for point sources and the instrumental response.  All errors are $\pm
1 \sigma$ (68\% confidence interval).}
\setlength{\extrarowheight}{0.05cm}
\begin{tabular}{lr@{}lr@{}lcr@{}lc}
\hline
Cluster &\multicolumn{2}{c}{$S_{\mathrm{Raw}}$} 
&\multicolumn{2}{c}{$S_{\mathrm{SC}}$}
&Fraction  &\multicolumn{2}{c}{$\Delta T$} &$y_0$\\ 
&\multicolumn{2}{c}{(mJy)} &\multicolumn{2}{c}{(mJy)} &Measured
&\multicolumn{2}{c}{($\mu$K)} &$\times10^4$\\ 
\hline
A1835 &-2.05{}&$\pm0.46$ &-3.82{}&$\pm0.47$ &0.363 
&-4050{}&$\pm500$ &7.70$\pm$0.95\\ 
ZW1953 &-1.65{}&$\pm0.37$ &-1.65{}&$\pm0.37$ &0.404 
&-1570{}&$\pm350$ &2.98$\pm$0.67 \\
%
%
ZW3146 &-1.06{}&$\pm0.39$ &-0.78{}&$\pm0.41$ & 0.181
&-1660{}&$\pm870$ &3.15$\pm$1.65\\
RX1532.9+3021 &-1.41{}&$\pm0.33$ &-1.11{}&$\pm0.34$ &0.166
&-2540{}&$\pm790$ &4.83$\pm$1.50\\
A2390 &-0.78{}&$\pm0.20$ &-1.02{}&$\pm0.22$ &0.176
&-2230{}&$\pm480$ &4.24$\pm$0.91\\
A2219 &11.10{}&$\pm0.38$ &-2.76{}&$\pm0.40$ &0.379
&-2810{}&$\pm400$ &5.34$\pm$0.76\\
RXJ2129.6+0005 &-0.13{}&$\pm0.37$ &-1.46{}&$\pm0.37$ &0.329
&-1710{}&$\pm430$ &3.25$\pm$0.82\\
A2261 &-3.19{}&$\pm0.30$ &-1.35{}&$\pm0.32$ &0.321
&-1610{}&$\pm390$ &3.06$\pm$0.74\\
A781  &0.68{}&$\pm0.42$ &-1.19{}&$\pm0.53$ &0.330
&-1390{}&$\pm620$ &2.64$\pm$1.18\\
A697 &-3.01{}&$\pm0.46$ &-3.01{}&$\pm0.46$ &0.422
&-2740{}&$\pm420$ &5.21$\pm$0.80\\
A1763 &10.58{}&$\pm0.80$ &-3.11{}&$\pm0.82$ &0.296
&-4040{}&$\pm1070$ &7.68$\pm$2.03\\
A68 &-0.89{}&$\pm0.48$ &-1.22{}&$\pm0.48$ &0.451
&-1040{}&$\pm410$ &1.98$\pm$0.78\\
A520 &-1.97{}&$\pm0.47$ &-1.80{}&$\pm0.48$ &0.268
&-2580{}&$\pm680$ &4.90$\pm$1.29\\
A267 &-2.73{}&$\pm0.43$ &-1.99{}&$\pm0.57$ &0.471
&-1620{}&$\pm470$ &3.08$\pm$0.89\\
RX0439.0+0715 &-1.22{}&$\pm0.43$ &-1.27{}&$\pm0.43$ &0.351
&-1390{}&$\pm470$ &2.64$\pm0.89$\\
ZWCL7160 &-1.10{}&$\pm0.34$ &-1.14{}&$\pm0.34$ &0.188
&-2330{}&$\pm700$ &4.43$\pm$1.33\\
A773 &-1.20{}&$\pm0.32$ &-1.21{}&$\pm0.32$ &0.382
&-1210{}&$\pm330$ &2.30$\pm$0.63\\
\hline
\end{tabular}
\label{tab:res}
\end{table*}

\subsection{SZ--X-ray scaling relations}
\label{sec:scaling}

According to the self-similar model of galaxy cluster formation, which
assumes evolution solely under gravity, we expect simple scalings
between various cluster parameters.  Following \cite{Morandi2007} we
derive the following relations, each of which may be tested using OCRA
data.  We note that while it is advantageous to use the integrated
Compton parameter for such studies (e.g. \citealp{Benson2004},
\citealp{Bonamente2008}), deriving this from OCRA data would introduce
further uncertainty since we have measured the cluster structures from
the X--ray data alone.  Thus, we restrict ourselves to discussing the
central Comptonisation, $y_0$. The scalings of interest are first, the
relation between central Comptonisation and X-ray temperature

\begin{equation}
y_0 \propto T_{\mathrm{X}}^{3/2} E(z),
\end{equation}

\noindent second, that between the Comptonisation and X-ray luminosity

\begin{equation}
y_0 \propto L_{\mathrm{X}}^{3/4} E(z)^{1/4},
\label{eq:yL}
\end{equation}

\noindent and finally, that between Comptonisation and
$Y_{\mathrm{X}}$, the product of gas mass and X-ray temperature

\begin{equation}
y_0 \propto Y_{\mathrm{X}}^{3/5} E(z)^{8/5},
\end{equation}

\noindent where $E(z)$ is as given in Equation \ref{eqn:E}.  Deviation
from the expected scalings would suggest contributions from
non-gravitational processes, and redshift dependence in the slope or
amplitude of the scalings should provide evidence on the nature of
these processes.

\begin{figure*}

\subfigure[X-ray temperature]{
\includegraphics[angle=270, width=0.57\linewidth]{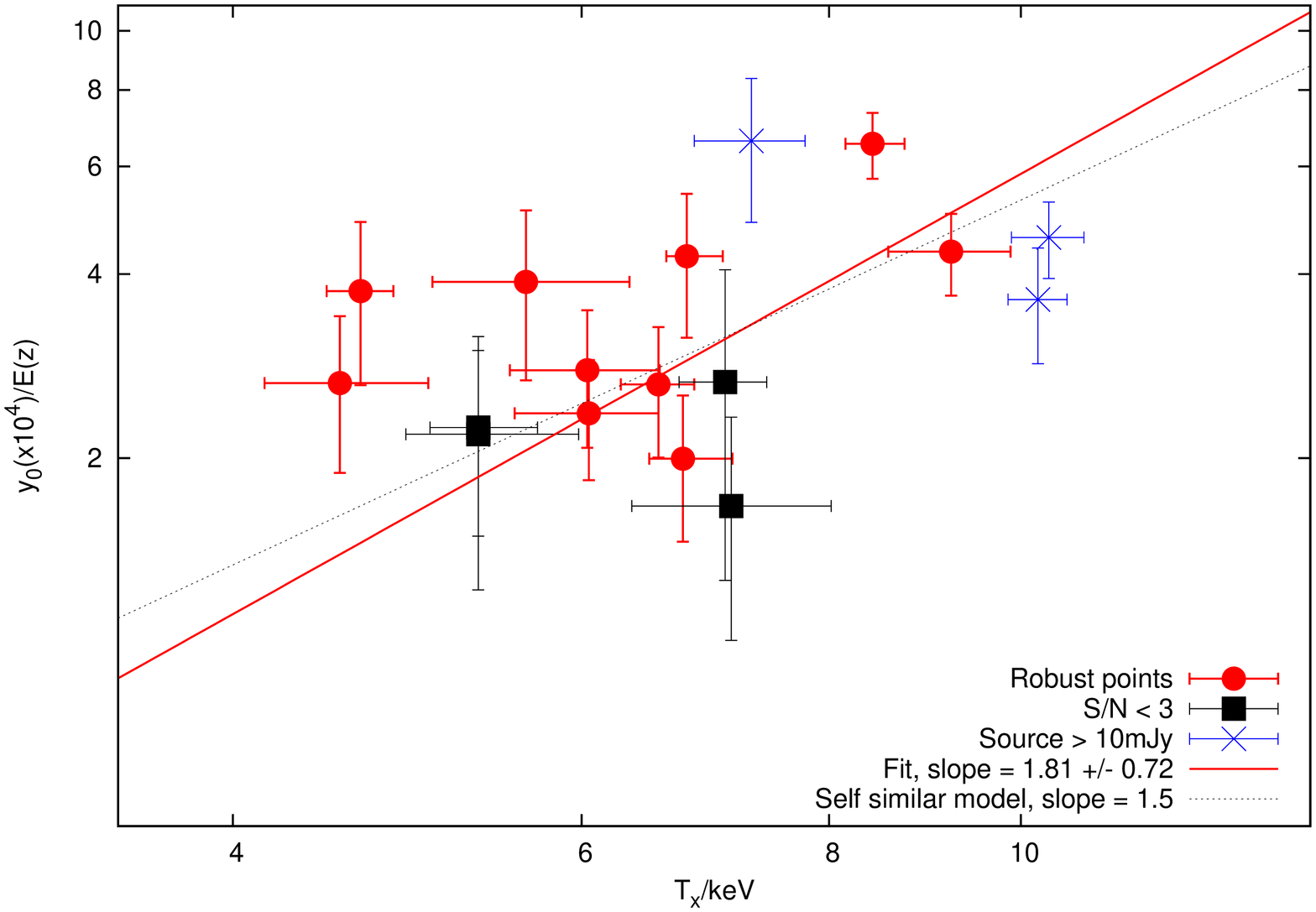}
\label{yT}
}
\subfigure[X-ray luminosity]{
\includegraphics[angle=270, width=0.57\linewidth]{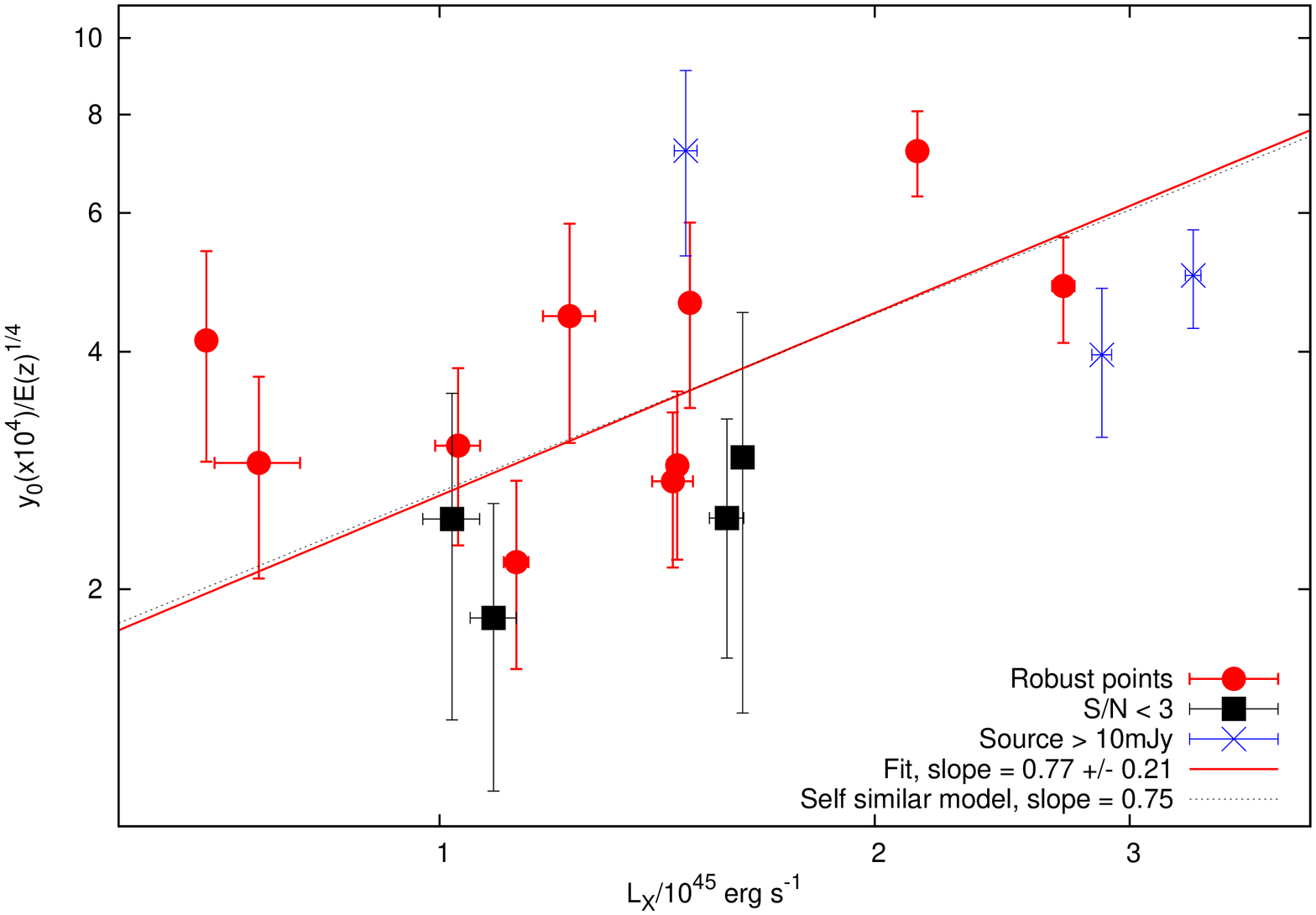}
\label{yL}  
}
\subfigure[$Y_{\mathrm{X}}$ parameter]{
\includegraphics[angle=270, width=0.57\linewidth]{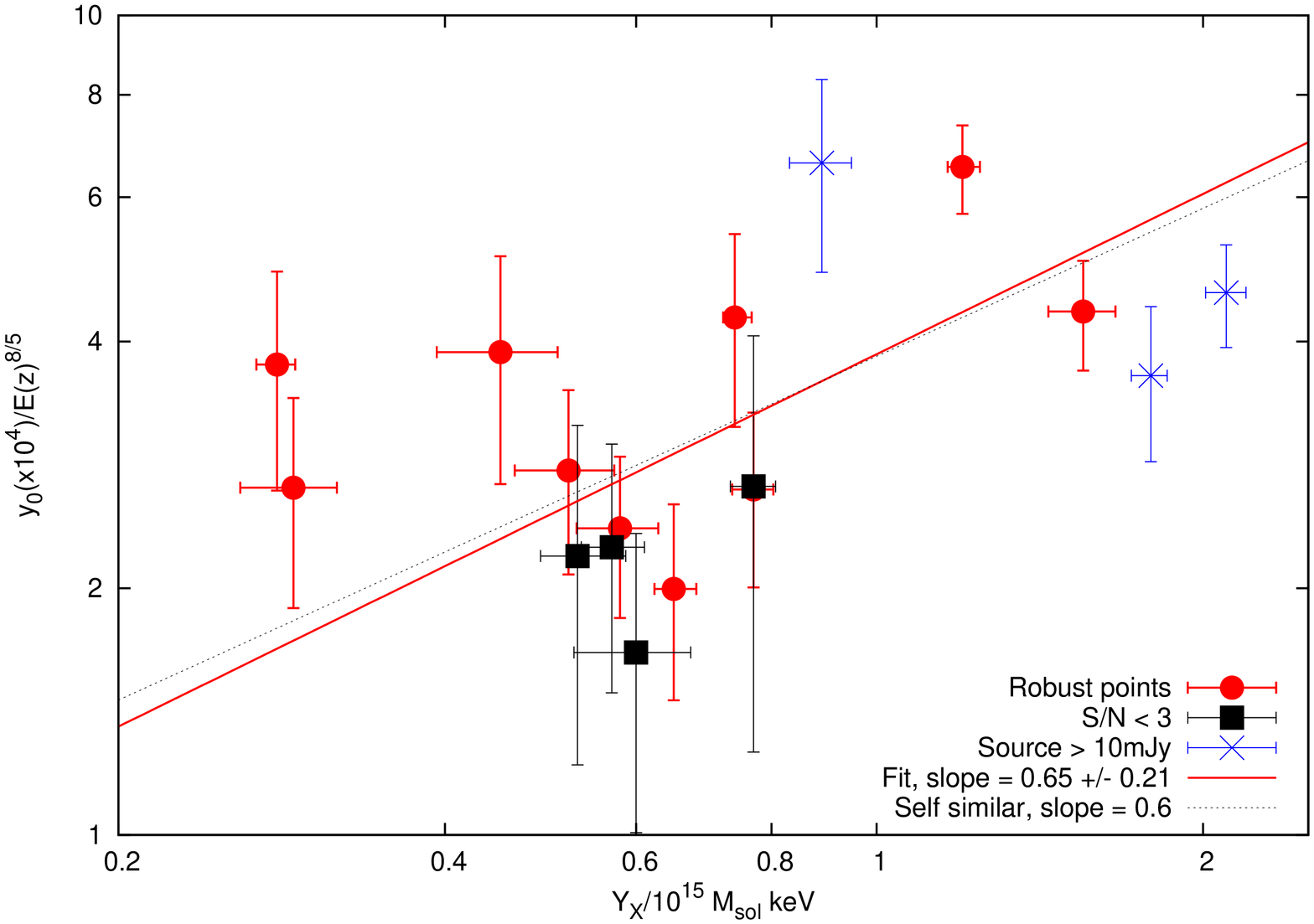}
\label{yY}  
}
\caption{Scaling relations between the central Comptonisation as
derived from OCRA data and various X-ray parameters.  We fit to the
`robust' and `low significance' data (red circles and black squares
respectively), but omit the clusters with potential residual source
contamination (blue crosses); the best--fit result is shown as the
solid red line.  The dotted black line depicts the expectation from
self--similar evolution.}
\end{figure*}

We present our scaling relations in figures \ref{yT} to \ref{yY}.  For
each plot, we colour--code the datapoints based on the reliability of
the OCRA data.  The most reliable points, i.e. those with no obvious
issues, are shown as red circles.  Points which have a signal to noise
ratio of less than 3 are black squares.  This is an arbitrary
cut--off, and we see no reason to exclude these points from our
analysis; however as an aid to the reader we choose to mark them
clearly.  Blue crosses represent clusters which may
suffer from residual source contamination, i.e. those which contain a
bright ($> 3$\,mJy) source, for which we have corrected, in their
central regions.  Due to the unavoidable 10 per cent uncertainty in the
calibration of the GBT data, we cannot be confident that the effects
of these sources have been accurately removed, and thus choose to
ignore these three clusters in our subsequent analyses.

\begin{table*}
\caption{Best--fit slopes for the correlations examined in Sections
\ref{sec:scaling} and \ref{sec:Xscale}, along with the expectations
from self--similar models for comparison. For the SZ/X-ray scalings,
we give results for the analysis both with and without corrections for
contaminant radio sources in the OCRA data. We note that the fits are
significantly poorer where the source corrections are omitted.  All
fits are consistent with predictions from self--similar models of
cluster formation. }
\begin{tabular}{c c l c c c}
\hline
Correlation &Expected slope &Data used &Best-fit slope &Fit statistic\\
\hline
$y_0, T_{\mathrm{X}}$ &1.5 &SZ, X-ray &$1.82\pm0.72$ &2.39\\
&&SZ (no source correction), X-ray &$1.94\pm1.10$ &6.64\\
\hdashline
$y_0, L_{\mathrm{X}}$ &0.75 &SZ, X-ray &$0.77\pm0.21$ &2.46\\
&&SZ (no source correction), X-ray &$0.71\pm0.19$ &6.86\\ 
\hdashline
$y_0, Y_{\mathrm{X}}$ &0.6 &SZ, X-ray &$0.65\pm0.21$ &2.50\\
&&SZ (no source correction), X-ray &$0.55\pm0.21$ &7.23\\
\hdashline
$L_{\mathrm{X}}, T_{\mathrm{X}}$ &0.5 &X-ray only &$0.49\pm0.04$ &3.40\\
$Y_{\mathrm{X}}, T_{\mathrm{X}}$ &2.5 &X-ray only &$2.19\pm0.16$ &1.85\\
$Y_{\mathrm{X}}, L_{\mathrm{X}}$ &1.25 &X-ray only &$1.24\pm0.05$ &4.29\\
\hline
\label{tab:summary}
\end{tabular}
\end{table*}

The fitting was performed in the linear space of observables, rather
than log--log space, and involved the minimisation of the squared and
error--weighted residuals from a power law of the form
$y=Ax^{\mathrm{B}}$, where $A$ and $B$ are to be estimated.  The space
of acceptable values of $A$ and $B$ was sampled by taking $10^5$
datapoint realisations, distributed according to their errors, assumed
Gaussian.  The best--fit slopes are summarised in Table
\ref{tab:summary}, with the values of the fit statistic.  Figure
\ref{yT} shows the scaling between $y_0$ and $T_{\mathrm{X}}$. For this
relation, the space of acceptable $(A,B)$ parameters extends over a wide
range in both parameters, so that the slope should be regarded as
essentially unconstrained, although the best fit value $B = 1.82\pm0.72$
is in agreement with the predicted value $B = 1.5$ from self--similar
models.  The other two relations show modest covariances between $A$ and
$B$ and are more robust.  Figure \ref{yL} shows the scaling between
$y_0$ and $L_{\mathrm{X}}$.  We obtain a slope of $0.77\pm0.21$ in good
agreement with self--similarity, for which Equation \ref{eq:yL} predicts
0.75.  Finally, Figure \ref{yY} depicts the scaling between $y_0$ and
$Y_{\mathrm{X}}$.  We derive a slope of $0.65\pm0.21$ which is in good
agreement with the predicted $B = 0.6$. For all three cases, the
intrinsic scatter in $y_0$ can be estimated by the additional error
required to bring the fit statistic down to a statistically acceptable
level $\sim1.0$ We find that the intrinsic scatter in $y_0$ is about 25
per cent. The errors on $B$ are large for all fits because of the modest
range of cluster masses in our sample, but we see no evidence for
departure from self--similar models of cluster evolution.  Our results
for the $y_0/L_{\mathrm{X}}$ and $y_0/T_{\mathrm{X}}$ scalings are
consistent with those obtained by \cite{Morandi2007}, although their gas
masses and temperatures were defined over significantly different
regions of the clusters (within $R_{2500}$), and their scalings have
smaller slope errors because of the wider range of $T_{\mathrm{X}}$ of
their clusters.

It is interesting to compare our scalings with those that we would
have obtained if source corrections were not available to us, since
blank--sky surveys for cluster SZ effects could attempt to constrain
the scalings with inadequate source data.  We find that the scatter
induced by the radio sources degrades the fit quality for each
correlation significantly, as illustrated by the fit statistics shown
in Table \ref{tab:summary}. Although the results broadly agree with
self--similar predictions, the residual contamination tends to cause
the scalings to appear too flat. For instance, in the high signal to
noise sub--sample where the effect of sources is clearest, the slope
of the $y_0/L_\mathrm{X}$ relation is found to be flatter by 0.27 if a
fit is attempted before source corrections are made. A similar result
is found for the $y_0/Y_{\mathrm{X}}$ scaling relation.

\subsection{X-ray only scaling relations}
\label{sec:Xscale}

We check the consistency of the set of clusters with the more usual
X--ray scaling relations between $L_{\mathrm{X}}$ and
$T_{\mathrm{X}}$, $Y_{\mathrm{X}}$ and $T_{\mathrm{X}}$, and
$Y_{\mathrm{X}}$ and $L_{\mathrm{X}}$ (e.g. \citealp{Morandi2007}) in
the same way as for the SZ / X--ray scaling relations (results also
summarised in Table \ref{tab:summary}).  The slopes that we measure,
of $0.49 \pm 0.04$, $2.19 \pm 0.16$, and $1.24 \pm 0.05$, are
consistent with the similarity expectations of 0.5, 2.5 and 1.25.
As the slight discrepancy for the correlation between $Y_{\mathrm{X}}$
and $T_{\mathrm{X}}$ is not of high statistical significance, this study
shows our sample to be representative of the population of hot ($T
\gtrsim 6$\,keV) clusters.

\section{DISCUSSION AND CONCLUSION}

We have observed a complete sample of galaxy clusters using OCRA--p,
and studied the scaling of the central Compton parameter, $y_0$, with
various X--ray quantities.  For each relation, we find slopes in good
agreement with the predictions from self--similar models.

Our study has similarities to that of \cite{Morandi2007}, and the two
samples have 7 clusters in common, although we imposed stricter
initial selection criteria.  \citeauthor{Morandi2007} consider
scalings with $y_0$, so we are able to perform a direct comparison.
They find that the $y_0/T_{\mathrm{X}}$ relation deviates by
$\sim3\sigma$ from the self--similar prediction, in the sense of being
steeper.  While our data are consistent with this result, we note that
they are essentially unable to constrain this relation at any level.
Regarding the $y_0/L_{\mathrm{X}}$ scaling, they find a slope of
$0.61\pm0.05$, which is in good agreement with our work. Thus we see
no sign of the flattening that would have resulted from the presence
of an undetected set of contaminating sources (Section
\ref{sec:scaling}).

More recently, \cite{Bonamente2008} studied a sample of 38
clusters observed with OVRO/BIMA and \emph{Chandra}.  They considered
the integrated Comptonisation, $Y$, and in particular its scaling with
$T_{\mathrm{X}}$, $M_{\mathrm{tot}}$ and $M_{\mathrm{gas}}$.  They
find good consistency with predictions from self--similar models for
all scaling relations.

Our analysis has been limited by the narrow region of parameter space
sampled, particularly for the scalings with $T_{\mathrm{X}}$.  Based
on our results, a sample double the size would provide a good test of
the $y_0/Y_{\mathrm{X}}$ scaling; this will be presented in an
upcoming paper in which we extend our sample to 33 clusters. To fully
test the other relations, a sample four times larger is required.
Thinking forwards to the prospects for blind cluster surveys, it is
interesting to note that in this study we would have detected Sunyaev
Zel'dovich effects for 14 out of 17 clusters in our sample with no
source information.  Of course, to accurately determine the strength
of the SZ effect in each case, and indeed to derive any cluster
parameters, good knowledge of the radio source environment is
essential, and any attempt to test the scaling relations without
taking sources into account will fail (see Section \ref{sec:scaling}).
That this is true for sub--mm sources, as well as cm--wave studies, is
clear from the large population of sub--mm sources known to be lensed
by massive clusters (e.g. \citealp{Johansson2010}).

\section*{ACKNOWLEDGMENTS}

We acknowledge support for the design and construction of OCRA-p from
the Royal Society Paul Instrument Fund, and funds for the data
acquisition system and operation on the telescope from the Ministry of
Science in Poland via grant number N N203 39043 who, along with STFC,
also supported the scientific exploitation of the completed system.
The GBT is a National Radio Astronomy Observatory (NRAO) instrument, a
facility of the National Science Foundation operated under cooperative
agreement by Associated Universities, Inc. We extend special thanks to
all the staff at the GBT, particularly Brian Mason and Carl Bignell,
for their essential support and prompt re--scheduling.


\bibliographystyle{mn2e}
\bibliography{ref_lancaster}


\bsp 

\label{lastpage}

\end{document}